\documentclass[aps,prb,twocolumn,showpacs,amssymb]{revtex4}
\usepackage{bm}
\usepackage{graphicx}

\begin{document}

\title{In-plane magnetic field phase diagram of superconducting $Sr_2RuO_4$}

\author{V.P. Mineev } 
\affiliation{Commissariat \`a l'Energie Atomique,
DSM/DRFMC/SPSMS 38054 Grenoble, France}

\date{\today}

\begin{abstract}
We develop the Ginzburg - Landau theory of the upper critical field  in the basal plane of a tetragonal multiband metal in two-component superconducting state. It is shown that typical for the two component superconducting state the upper critical field basal plane anisotropy and the phase transition splitting still exist in a multiband case. However, the value of anisotropy 
can be effectively smaller than in the single band case.  The results are discussed in the application to the superconducting $Sr_2RuO_4$.

\end{abstract}

\pacs{74.20.De, 74.20.Rp, 74.25.Dw, 74.70.Pq}

\maketitle

\section{Introduction}

The tetragonal compound $Sr_2RuO_4$ is an unconventional superconductor (see review \cite{Mack}). 
It reveals properties typical for non-s-wave Cooper pairing: the suppression of superconducting state by disorder \cite{Mac}, the presence of 
zeros in the superconducting gap discovered by the magnetothermalconductivity measurements \cite{ Tanatar,Izawa},  the odd parity of the superconducting state in respect of reflections in $(a,c)$ plane  established by the Josephson interferometry method \cite{Liu}.
All these properties  are equally possible for single or multi component order parameter superconducting state.

Other important observations demonstrate the appearance of spontaneous magnetization or  time reversal symmetry breaking in the superconducting state of this material. These are: (i)  the increasing  of $\mu$SR  zero-field relaxation rate 
\cite{Luke}, (ii) the hysteresis observed in field sweeps of the critical  Josephson current \cite{Kid}  and (iii) the Kerr rotation of the polarization direction of reflected light from the surface of a superconductor \cite{Xia} (for the theoretical treatment see \cite{Yak, Min1}).

A superconducting state  possessing  spontaneous magnetization is described by multicomponent order parameter \cite{Min}. In a tetragonal crystal the  superconducting states with two-component order parameters $(\eta_x,\eta_y)$ corresponding  to singlet or to triplet pairing are admissible \cite{Min}. In application to $Sr_2RuO_4$ the time reversal symmetry breaking form of the order parameter 
$(\eta_x,\eta_y)\propto(1,i)$ has been proposed first in the paper \cite{Rice}.
The specific properties for the superconducting state with two-component order parameter  in a tetragonal crystal under magnetic field in basal plane are (i) the anisotropy of the upper critical field \cite{Gor,Bur} and (ii) the  splitting of the phase transition to superconducting state in two subsequent transitions \cite{Min}. 
Both of these properties should manifest themselves 
starting 
from the Ginzburg-Landau temperature region $T\approx  T_c$ but till now there is no experimental evidence for that. The in-plane  anisotropy of the upper critical field has been observed only at low temperatures \cite{Mao} where it  is quite well known phenomenon for any type of superconductivity
originating from the Fermi surface anisotropy.  
Theoretically in application to $Sr_2RuO_4$ these properties  have been investigated by Kaur, Agterberg, and Kusunose \cite{Kaur}.  They have found that 
one particular choice of  the basis functions of two-dimensional irreducible representation for a tetragonal
point group symmetry  is appropriate for elimination of basal plane upper critical field anisotropy but at the same time the considerable phase transition splitting occurs. 
Vice versa, another particular choice of the basis functions almost eliminates the phase transition splitting for the the one particular field direction but keeps the basal plane upper critical field anisotropy.
Thus the basal plane upper critical field properties look as incompatible with multicomponent order parameter structure dictated by the  experimental observations manifesting the spontaneous time-reversal breaking. 

All the mentioned theoretical treatments of $H_{c2}$ problem have been undertaken for the two component
superconducting state in a single band superconductor. On the other hand,  in $Sr_2RuO_4$ we deal with three bands of charge carriers \cite{Mack}. Hence, the formation of  multiband superconducting state is quite probable. A microscopic theory of such a state was  proposed in the paper \cite{Zhit}. 

Here, in the application to
the problem of the upper critical field in the basal plane of a tetragonal crystal we shall develop a phenomenological theory of multiband multicomponent superconducting state. It will be shown that both properties:  $H_{c2}$ basal plane anisotropy and the phase transition splitting still exist in a multiband case. However, quantitatively, the value of anisotropy 
can be smaller than in the single band case.  

\section{Upper critical field}

The order parameter in multiband tetragonal superconductor with singlet pairing is
\begin{equation}
\Delta({\bf r},\hat{\bf k})=
\sum_{\lambda}¥\sum_{i}\eta_{i}^\lambda¥({\bf
r})\psi_{i}^\lambda¥(\hat{\bf k}).
\label{ea}
\end{equation}
Here the lower latin index $i=x,y$ numerates the components of the order parameter,  the upper
greek index is the band number, and $\psi_{i}^\lambda¥(\hat{\bf k})$   
are the functions of the irreducible
representation  dimensionality $2$ of the point symmetry group
$D_{4h}$ of the crystal in the normal state.  Similar decomposition takes
place for vectorial order parameter function in triplet state
\begin{equation}
{\bf d}({\bf r},\hat{\bf k})= \sum_{\lambda}\sum_{i}\eta_{i}^\lambda¥({\bf
r})\mbox{\boldmath $\psi$}_{i}^\lambda¥(\hat{\bf k}).
\label{eb}
\end{equation}
Although our theory is applicable to the superconductor with arbitrary number of bands we shall write all the concrete results for the two band situation.

 Following derivation given in the paper \cite{Min2} one can easily obtain the generalization of the G-L equations for a two-component superconducting state in a tetragonal crystal \cite{Bur} for multiband case 
\begin{eqnarray}
&g^{\lambda\mu}&\left[K_1^\mu D_i^2\eta_j^\mu+
K_2^\mu D_jD_i\eta_i^\mu+K_3^\mu D_iD_j\eta_i^\mu+K_4^\mu D_z^2\eta_j^\mu
\right.\nonumber\\
&+&\left.
K_5^\mu(\delta_{xj}D_x^2\eta_x^\mu+\delta_{yj}D_y^2\eta_y^\mu) -
\Lambda(T)\eta_j^\mu \right ]+
\eta_j^\lambda=0.
\label{e1} 
\end{eqnarray}
Here $$D_i=-i\frac{\partial}{\partial r_i}+\frac{2e}{c}A_i({\bf r})$$ is the operator of covariant differentiation, the Planck constant $\hbar$ is taken equal to unity throughout the paper, and the function $\Lambda(T)$ is $$\Lambda=\ln\frac{2\gamma\epsilon}{\pi T},$$ where $\ln\gamma=0,577...$ is the Euler constant, $\epsilon$ is an energy cutoff for the pairing interaction.
We assume here that it has the same value for the different bands. The matrix $g^{\lambda\mu}$ is
$$
g^{\lambda\mu}=
V^{\lambda\mu}\langle|
\mbox{\boldmath $\psi$}_{i}^\mu¥(\hat{\bf k})|^{2}¥ N_{0}^\mu¥
(\hat{\bf k})\rangle,
$$
here $V^{\lambda\mu}$ is the matrix of the constants of pairing interaction. The angular brackets mean the
averaging over the Fermi surface, $N_{0}^\mu¥(\hat{\bf k})$ is
the angular dependent density of electronic states at the Fermi
surface of the band $\mu$.  The gradient terms coefficients are
$$
K_{1}^\mu=\frac{\langle|\mbox{\boldmath $\psi$}_{x}^\mu¥(\hat{\bf k})v_{Fy}^\mu¥(\hat{\bf k})|^2¥ N_{0}^\mu¥ (\hat{\bf
k})\rangle}{\langle|
\mbox{\boldmath $\psi$}_{i}^\mu¥(\hat{\bf k})|^{2}¥ N_{0}^\mu¥
(\hat{\bf k})\rangle}
\frac{\pi T}{2}\sum_{n\ge
0}¥\frac{1}{|\omega_{n}¥|^{3}¥},
$$
$$
K_{2}^\mu=\frac{\langle\mbox{\boldmath $\psi$}_{x}^\mu¥(\hat{\bf k})(
\mbox{\boldmath $\psi$}_{y}^\mu(\hat{\bf k}))^{*}
v_{Fx}^\mu¥(\hat{\bf k})v_{Fy}^\mu¥(\hat{\bf k}) N_{0}^\mu¥ (\hat{\bf
k})\rangle}{\langle|
\mbox{\boldmath $\psi$}_{i}^\mu¥(\hat{\bf k})|^{2}¥ N_{0}^\mu¥
(\hat{\bf k})\rangle}
\frac{\pi T}{2}\sum_{n\ge
0}¥\frac{1}{|\omega_{n}¥|^{3}¥},
$$
$$K_{3}^\mu=K_{2}^\mu,$$
$$
K_{4}^\mu=\frac{\langle|
\mbox{\boldmath $\psi$}_{x}^\mu¥(\hat{\bf k})
v_{Fz}^\mu¥(\hat{\bf k})|^2¥ N_{0}^\mu¥ (\hat{\bf
k})\rangle}{\langle|
\mbox{\boldmath $\psi$}_{i}^\mu¥(\hat{\bf k})|^{2}¥ N_{0}^\mu¥
(\hat{\bf k})\rangle}
\frac{\pi T}{2}\sum_{n\ge
0}¥\frac{1}{|\omega_{n}¥|^{3}¥},
$$
$$
K_{5}^\mu=\frac{\langle|\mbox{\boldmath $\psi$}_{x}^\mu¥(\hat{\bf k})
v_{Fx}^\mu¥(\hat{\bf k})|^2¥ N_{0}^\mu¥ (\hat{\bf
k})\rangle}{\langle|
\mbox{\boldmath $\psi$}_{i}^\mu¥(\hat{\bf k})|^{2}¥ N_{0}^\mu¥
(\hat{\bf k})\rangle}
\frac{\pi T}{2}\sum_{n\ge
0}¥\frac{1}{|\omega_{n}¥|^{3}¥},
$$
where $\omega_{n}=\pi T(2n+1)$ is the Matsubara frequency, and the components of the Fermi velocity of the band $\mu$ are given by 
$v_{Fx}^\mu¥(\hat{\bf k}), v_{Fy}^\mu¥(\hat{\bf k}),v_{Fz}^\mu¥(\hat{\bf k})$. The definition of $K^\mu$ coefficients accepted here differs  from the trational one \cite{Min2} by the terms in the denominators.

Neglecting the gradient terms and taking the determinant of the system (\ref{e1}) equal to zero we obtain the critical temperature
\begin{equation}
T_{c}¥=\frac{2\gamma\epsilon}{\pi}\exp{(-1/g)},
\label{e2}
\end{equation}
where $g$ is defined by 
\begin{equation}
g= (g^{11}¥+g^{22}¥)/{2} +\sqrt{({g^{11}¥-g^{22}¥})^{2}¥/{4}+g^{12}¥g^{21}¥}.
\label{e3}
\end{equation}
The matrix $g^{\lambda\mu}$ in tetragonal cristal has the common value for $x$ and $y$ components of the order parameter.  Hence the phase transition to superconducting state occurs at the same critical temperature for all the component of the order parameter in all the bands.

In the case of a magnetic field in the basal plane, $$ {\bf H}=H(\cos\varphi,\sin\varphi,0),$$
$${\bf A}=Hz(\sin\varphi,-\cos\varphi,0)$$  we obtain from (\ref{e1})
\begin{eqnarray}
&g^{\lambda\mu}&\left[-(K_4^\mu \partial_z +
\Lambda)\delta_{ij} \right.\nonumber\\
&+&\left.h^2z^2\left( \begin{array} {cccc}K_1^\mu+K_{235}^\mu\sin^2\varphi & -K_{23}^\mu\sin2\varphi\\
-K_{23}^\mu\sin2\varphi & K_1^\mu
+K_{235}^\mu\cos^2\varphi \end{array}
\right)_{ij}
\right ]\eta_j^\mu\nonumber\\
&+&
\eta_j^\lambda=0.
\label{e4} 
\end{eqnarray}
Here we have introduced notations $h=2\pi H/\Phi_0$, $K_{23}^\mu=K_2^\mu+K_3^\mu$, $K_{235}^\mu=K_2^\mu+K_3^\mu+K_5^\mu$.

Making use the orthogonal transformation
\begin{equation}
\tilde\eta_p^\mu=\left( \begin{array} {cccc}\cos\beta^\mu& \sin\beta^\mu\\
-\sin\beta^\mu&\cos\beta^\mu\end{array}
\right)_{pl}\eta_l^\mu,~~~\tan2\beta^\mu=\frac{K_{23}^\mu}{K_{235}^\mu}\tan2\varphi
\label{e5}
\end{equation}
we come to
\begin{equation}
g^{\lambda\mu}\left[-(K_4^\mu \partial_z +
\Lambda)\delta_{ij} 
+h^2z^2\left( \begin{array} {cccc}b_x^\mu& 0\\
0& b_y^\mu\end{array}
\right)_{ij}
\right ]\tilde\eta_j^\mu+
\tilde\eta_j^\lambda=0.
\label{e6} 
\end{equation}
Here
\begin{equation}
b_{x,y}^\mu=K_1^\mu+\frac{K_{235}^\mu\pm\sqrt{\left(K_{235}^\mu\cos2\varphi\right)^2+
\left(K_{2}^\mu\sin2\varphi\right)^2}}{2}
\label{e7}
\end{equation}
It follows from Eqn. (\ref{e6}) that in finite magnetic field the phase transition to superconducting state splits on two subsequent phase transitions. Indeed, the systems of equations for    $x$ or $y$  order parameter components are independent, hence they have independent and non-equal eigen values.
The corresponding upper critical fields  can be found only numerically. Here we solve this problem following a variational approach, which is known to give a good accuracy in similar cases \cite{Dao}.
So, we look for a solution for the $x$ component of the order parameter in the form
\begin{equation}
\tilde\eta_x^\mu=\left( \begin{array} {cccc}\tilde\eta_x^1\\
\tilde\eta_x^2\end{array}\right)=\left(\frac{\lambda_x}{\pi}\right)^{1/4}
\left( \begin{array} {cccc}C_x^1\\
C_x^2\end{array}\right)e^{-\frac{\lambda_xz^2}{2}}.
\label{e8}
\end{equation}
The similar formula and the following calculations are valid for the $y$-component of the order parameter.

After substitution of (\ref{e8}) in Eqn. (\ref{e6}), multiplication of it by $\exp(-\lambda_xz^2/2)$  and spacial integration we obtain
\begin{equation}
g^{\lambda\mu}(E_x^\mu-\Lambda)C_x^\mu +C_x^\mu=0,
\label{e9}
\end{equation}
where
\begin{equation}
E_x^\mu=\frac{K_4^\mu\lambda_x^2+h^2b_x^\mu}{2\lambda_x}.
\label{e10}
\end{equation}
The transition field is determined by condition of vanishing of the determinant of the system (\ref{e9}).
Particularly we are interested in upper critical field behavior near the critical temperature.
Obviously, $E_x^\mu\propto h$, hence it tends to zero at $T\to T_c$. So, in vicinity of critical temperature 
we receive after the simple calculations
\begin{equation}
\ln\frac{T_c}{T}=\frac{E_x^1(1+a)+E_x^2(1-a)}{2},
\label{e11}
\end{equation}
where $T_c$ is determined by Eqn. (\ref{e2}) and
\begin{equation}
a=\frac{g^{11}-g^{22}}{\sqrt{(g^{11}-g^{22})^2+4g^{12}g^{21}}}.
\label{e12}
\end{equation}
The maximum of critical temperature at non-zero magnetic field is accomplished at following $\lambda_x$
value
\begin{equation}
\lambda_x|_{max}=h\lambda_x^0,~~~~\lambda_x^0=\sqrt{\frac{\tilde b_x}{\tilde K_4}},
\label{e13}
\end{equation}
where
\begin{equation}
\tilde b_x=b_x^1(1+a)+b_x^2(1-a),
\label{e14}
\end{equation}
\begin{equation}
\tilde K_4=K_4^1(1+a)+K_4^2(1-a).
\label{e15}
\end{equation}
So, after the substitution of  Eqns. (\ref{e10}), (\ref{e13}), (\ref{e14}), (\ref{e15})  into Eqn. (\ref{e11}) we obtain
\begin{equation}
h_{x,y}=\frac{2(1-T/T_c)}{\sqrt{\tilde K_4\tilde b_{x,y}}}.
\label{e16}
\end{equation}

In the case of single band superconductivity our variational solution is exact  and we obtain from 
(\ref{e15}) droping out all the terms with index $\mu=2$
\begin{equation}
h_{x,y}=
\frac{1-T/T_c}{\sqrt{b_{x,y}^1K_4^1}}.
\label{e17}
\end{equation}

So, the situation for the two-band and one-band superconducting states is characterized by
the same properties: the  basal plane anisotropy of the upper critical field corresponding to the largest of two eigenvalues $h_y$ and 
two consecutive phase transitions to the superconducting state, first with  $y$ component and when with $x$ and $y$ components of the order parameter. 

It is worth noting that in the multiband case due to the compensation of the different bands contribution \cite{Agt2001} the actual value of the anisotropy of $H_{c2}$ can be smaller than in one-band situation.
The phase transition splitting, however,  still  persists in multiband case. The absence of experimental evidence of this phenomenon argues in support of one component superconducting state in 
$Sr_2RuO4$. 

\section{Conclusion}

 We have demonstrated that for  a two-component superconducting state in a tetragonal crystal the basal plane anisotropy of the upper critical field and the phase transition splitting are inherent properties both
for the one band and multiband superconductivity.
Thus, the experimentally established \cite{Mao} absence of these phenomena  in $Sr_2RuO_4$ says 
opposite  to the possibility of existence of two component superconducting state in this material.

\section*{ACKNOWLEDGEMENTS}

I am indebted to M.Zhitomirsky, Y. Liu and  D.Agterberg for the discussions  of the problem
of upper critical field in the mlticomponent multiband superconducting state.

\end{document}